# Low–Temperature Plasma Technology: Impact on Indian Rural Life

Mangilal Choudhary

Institute of Advanced Research, The University for Innovation, Koba, Gandhinagar, 382426, India

**ABSTRACT**

The low temperature plasma technology has been emerged as an alternative environmental friendly technology that has a potential to replace the traditional technologies in various sectors such as textile industries, agriculture, food industries and waste water treatment. The future prospective of the low temperature plasma technology on the life of rural Indians has been discussed in this brief report.

## I. INTRODUCTION

What are people's most common thought about the states of matter existing in the universe? Are these solid, liquid and gas only? The answer is no but why? Does any other state of matters exist? Yes but which state? The fourth state of matter, named as 'PLASMA'. It is an ionized gas consists of negatively charged electrons and positively charged ions in which these charged particles interact via long range coulomb interaction and capable to exhibit the collective response to an external field force [1]. Depending on the averaged energy of the electrons and ions in the mixture of ionized gas (plasma), it can be named as hot plasma (or thermal plasma) and cold plasma (or non-thermal plasma). Some examples of the hot plasma are thermonuclear fusion plasma (e.g. sun), atmospheric arcs, flames and sparks. In a plasma, the average energy of ions is negligible than the electrons. Examples of cold plasmas include the discharge in a fluorescent tube, different glow discharges (dielectric barrier discharge, surface micro discharge) in laboratory, and Earth's ionosphere [2-4]. In laboratory, non-thermal plasma is produced by the electric breakdown of gases at atmospheric pressure. A high voltage difference across the electrodes is required to breakdown the gases at atmospheric pressure. In the non-thermal plasma, the majority of the discharge energy goes into the production of energetic electrons, rather than ion and neutron heating. These energetic electrons are main source for production of radicals or reactive species of nitrogen and oxygen (RONS) through the electron impact dissociation and ionization processes [5, 6]. The role of the various reactive species, UV radiation, energetic electrons, and ions of non-thermal plasmas has been studied in multidisciplinary scientific fields, which are discussed in subsequent section.

## II. TEXTILE INDUSTRIES

The Non-thermal plasma (NTP) plasma has been widely using in the textile industries worldwide. The textile materials in the form of sheets, fabrics and polymers are treated with the non-thermal atmospheric pressure plasma to modify the surface properties and making surface anti-bacterial. In treating with the plasma, the modified surface wettability and surface textures may increase dye or finishing agents' absorptions, surface chemical and morphological Modifications help to improve the hydrophilicity of fabrics, and reduction of bacterial species on the fabric surface [7–9]. There are chemical processes to treat the wools and cottons fabrics to modify the surface properties of raw materials to make them into consumable products. The use of NTP technology to treat the wools and cotton fabrics has some advantages over the other processes. It is possible to treat all the organic compounds by NTP, it only modifies the surface properties without altering the bulk characteristics and it is an environment friendly and required a less water usage [7, 8].

In rural India, sheep and goats play an important role in livehood of a large proportion of small farmers and landless labourers. The production of wool from sheep and goats is a part of the source of income. The wool quality of mostly sheep and goats of Indian states is not satisfactory, which is a main cause of the low demand and limited production. Due to the less demand of Indian sheep wool because of its quality, it does not contribute more to the net income of sheep farming sector. Improvement of quality of wools is one of the factors to make its value in market which definitely helps to increase the net profit of the sheep farmers. The non-thermal plasma technology has a potential to improve the quality of the wool fibres at low input costs [7, 9].

Apart from wools, the cotton is also a major crop in India. The cotton fibres are most important textile fibres and very comfortable. In some parts of country (rural region), it is main crop for the source of income for farmers. The improved quality of cotton fibres brings the most money in the commodity markets. Instead of traditional method to improve the quality of cotton fibres, plasma can also modify the surface of cotton fibres. Hence, plasma technology is considered as an advanced tool to improve the quality of the wools and cotton fibres [7–9]. In this view, there is need of installation of small scale industries of textiles in the rural region for converting row material (wools and cotton) to finished/usable products. The low energy consumable plasma technology can be a part of the surface modification (or quality improvement) unit in the textile industries. There is a possibility to use of different types of plasma sources such as dielectric barrier discharge plasma, energetic electrons beam, and ions beam to modify the surface properties of fibres. For reducing the input cost, it is also possible to operate these systems with the help of solar powers. Hence the low energy cost, less water usage, and environment friendly non-thermal plasma technology is necessary to increase the cost of raw materials, which definitely will help to increase the net income of small farmers, marginal farmers and landless labourers who are involved in sheep farming sector.

### III. AGRICULTURE SECTOR

Agriculture is a back bone of the Indian economy. Indian's most of agriculture land is used for farming many crops and for growing vegetables and fruits. However, yields per hectare of crops in India are generally low compared to international standards. There are many factors such as shortage of water, fewer nutrients in soils, lack of suitable fertilizers, harmful insects and bugs etc. which strongly affect the yield of the crops. The low yield of crops/vegetables/fruits is the main cause of the low income of small and marginal farmers in rural region. It is true that India's population growth rate is positive which clearly indicates the increase in food demand for coming years. Therefore, obtaining high yield in agriculture production is necessary to compensate the increased food demand.

In recent years, low temperature plasma or non-thermal plasma is considered as an advanced green technology for enhancing productivity in agriculture sectors. The low temperature plasma has a potential to boost yield in a robust way without demanding more water and fertilizers. In this technology, crops, seeds and soil are treated to see the combined effects on the productivity [10, 11]. The low temperature plasma or plasma treated water helps in enhancing seeds germination, increasing the rooting speed, stimulating plant growth, preventing the pests and bugs, inactivating of micro-organism and fungal spores etc. The presence of reactive species of nitrogen in plasma along with water can form the nitrogen rich fertilizer which is essential nutrient for growth of the crops and keep them healthy. The presence of ozone, oxygen reactive species and UV radiation in non-thermal plasma can help to inactivate the microorganism, pests and fungal spores to keep the crops healthy during the entire life cycle. The productivity in agriculture sectors can be improved by keeping crops/plants healthy which is possible by using the non-thermal atmospheric-pressure plasma sources such as DBD, plasma jets and surface micro discharges [10–12].

Indian government is more focused on the doubling the income of farmers in coming years. It is only possible by increasing yield of crops per hectare with low input costs. Providing high quality seeds, low cost pesticides, and low cost fertilizers to farmers are essential to reduce the input cost of farming and increase the productivity. Day by day use of chemicals (fertilizers and pesticides) in the agriculture sectors is responsible for the poor health of human beings. To keep the society or people's healthy; alternate advanced eco-friendly technologies for increasing productivity of food grains, vegetables and fruits are demand for the future. It has been experimentally proven in many research labs worldwide that low temperature plasmas (or NTP) are an alternate green technology to boost the agriculture sector [10,11]. It is necessary to establish a broad multidisciplinary research network (plasma, agriculture, biotechnology and chemistry) among different research institutes/universities of country. Apart from the research activity, applied research projects with industrial partners are necessary to scale up the small laboratory projects at large scale for direct applications of non-thermal plasma technology to crops/plants in the rural region. It is expected to increase the productivity up to 20 % by using the low temperature plasma technology throughout the entire life cycle (from seed germination to inactivation of micro-organism) of crops/plants. The higher productivity will increase the net income of small and marginal farmers which is essential to survive a better life in the future.

**IV. FOOD SECTOR**

In India, the storage of vegetables/fruits for a long time is one of the big challenges. Due to the lack of cold storage chain between farmers and consumers, small and marginal farmers unable to sell the agriculture products such as vegetables and fruits on the reasonable price. A used amount of vegetables and fruits is wasted due to the growth of microbes under the suitable environmental conditions. However, there are some commonly used chemical agents

(organics acids) and radiations for inactivation of microbes on the vegetables and fruits. But the non-thermal plasma technology could be an alternative to presently used methods without any side effects. It has been reported in many studies that non-thermal atmospheric pressure plasma consists of different reactive oxygen and nitrogen species (RONS), energetic electrons, UV radiation and ions. Therefore, it has a potential to inactive different contaminating microbes (or surface disinfect) on vegetables and fruits [12–14]. The life time or freshness of vegetables/fruits is increased after treating with the non-thermal atmospheric pressure plasma.

It is possible to treat either directly with gas plasma or plasma activated water to decontaminate vegetables and fruits [13, 15]. Hence plasma treatment could be used to preserve the vegetables/fruits freshness and all other properties for few days (2 to 5 days) without cold storage. This small time period (2 to 5 days) is sufficient to transport the vegetable/fruits from the local region (or farm houses) to metro-city cold storage or consumers. The sterilization of packing materials (plastics bags/container) using atmospheric pressure plasmas also increase the life time of vegetables/fruits [12–14]. In India, solar power is assumed to be a future energy source. Similar to solar water pump at farm houses in rural region, solar power driven low temperature plasma sources could be established at village level for treating the vegetables/fruits and packing materials before transporting to other cities. Thus, the low cost and environment friendly plasma technology can help to increase the net income of small and marginal farmers of the rural India.

## V. WATER PURIFICATION

In rural India, hand pumps, tube wells and normal wells (ground water) are mainly sources of drinking water. The quality of ground water sometimes is not good. The bacterial contamination of pumping out water continues to be a widespread problem in the rural region of India. The poor quality of drinking water due to bacterial and viral species is a major cause of illness and death affected by water-borne diseases. Such water related diseases put a financial burden on low income sections in villages. Providing safe drinking water to all homes in rural India is a very challenging task. There are many technologies to make water drinkable but also require some advanced technologies. There is an emerging technology based on plasma (plasma-based water purification) having potential to inactive/remove the micropollutantsm, bacterial species and viruses from the pumping out water [16]. The interaction of plasma with water generates various reactive species (short and long lived) that inactive/kill the bacterial contamination of water [17, 18]. For providing the safe water to all the villagers of rural regions of India, we should use the low cost advanced water purification techniques. However, there are many technical challenges to install the plasma-based water purification reactors at local level but this could be a success mission in the coming years.

## VI. SUMMARY

In this report, a brief introduction about plasma and types of plasmas based on average energy of charged species (electrons and ions) are presented. The plasma with lower ions temperature than the electrons temperature is named as low temperature or non-thermal plasma. Applications of non-thermal plasmas in various fields such as textiles industries, agriculture, food sector, and waste water treatment are discussed. How an environment friendly plasma technology could transform the life of rural regions of India in coming years and challenges in implementing the technology at local level have been thoughtfully discussed.